\documentclass[twocolumn,aps,prd,showpacs]{revtex4}
\usepackage{graphicx}
\usepackage{latexsym}
\usepackage{amsmath}
\usepackage{amsfonts}
\usepackage{amssymb}
\usepackage{verbatim}

\newcommand{\be}{\begin{equation}}
\newcommand{\ee}{\end{equation}}
\newcommand{\bea}{\begin{eqnarray}}
\newcommand{\eea}{\end{eqnarray}}

\newcommand{\pa}{\partial}

\newcommand{\bb}{\bibitem}
\def\pls{\partial\!\!\!/}
\def\bb{\bibitem}
\def\as{A\!\!\!/}

\def\ps{p\!\!\!/}
\def\bs{b\!\!\!/}

\def\bb{\bibitem}
\newcommand{\ben}{\begin{eqnarray}}
\newcommand{\een}{\end{eqnarray}}

\begin{document}

\title{Lorentz-violating Chern-Simons action under high temperature in massless QED}
\author{$^{a}$F. A. Brito, $^{b}$L. S. Grigorio, $^{c}$M. S. Guimaraes, $^{a}$E. Passos,
$^{b}$C. Wotzasek} \affiliation{$^{a}$Departamento de F\'\i sica,
Universidade Federal de Campina Grande, Caixa Postal 10071,
58109-970  Campina Grande, Para\'\i ba,
Brazil\\
$^{b}$Instituto de F\'\i sica, Universidade Federal do Rio de Janeiro,
\\Caixa Postal 21945, Rio de Janeiro, Brazil
\\
$^{c}$Departamento de F\'\i sica Te\'orica, Instituto de F\'\i sica, UERJ - Universidade do Estado do Rio de Janeiro, Rua S\~ ao Francisco Xavier 524, 20550-013 Maracan\~ a, Rio de Janeiro, Brasil
}


\begin{abstract} 

Lorentz and CPT violating QED with massless fermions at finite temperature is studied. We show that there is no ambiguity in the induced coefficient of the Chern-Simons-like term that defines the so-called Carroll-Field-Jackiw model at high temperature. We also show that this system constitutes an example where the breaking of CPT and Lorentz symmetries is more severe at high temperature than in the zero temperature case thus precluding any naive expectations of Lorentz symmetry restoration.

\end{abstract}
 \maketitle


\section{Introduction}

There is wide interest nowadays in model field theories that provide phenomenological description of possible Lorentz and CPT symmetries violations. It has thus become increasingly important to address the question of stability of these symmetry violations as a function of environmental variations. It is expected that Lorentz and CPT violations are spontaneously broken, meaning that it is  a property of the particular state in which the system rests at low energies. It is then conceivable that symmetry restoration takes place at high temperatures where the temperature is responsible for setting the energy scale. But the restoration of symmetries at high temperatures is a naive lore known not to be true in general \cite{ns}.

The main purpose of the present work is to discuss a field theory model where not only the Lorentz symmetry is not restored but in fact the breaking is enhanced as the temperature gets higher.  We will address the specific problem of the radiatively induced Lorentz and CPT violating term in QED at finite temperature. This modified electrodynamics is known as the Carroll-Field-Jackiw (CFJ) model \cite{cfj} and can be seen as a sector of the Standard Model Extension (SME) \cite{emp}.

There has been conflicting reports in the literature about the induction of the Chern-Simons-like term defining the CFJ model in the finite temperature framework of massive QED \cite{Cervi,Ebert,N3,bpp,gnpp}. Different approaches seems to lead to different conclusions about the value of the numerical coefficient of this term. For example, in the reference \cite{Cervi}, the authors argue that at finite temperature the Chern-Simons-like term is generically present, with the value of its coefficient being unambiguously determined up to a temperature independent constant, related to the zero temperature renormalization conditions. The authors of reference \cite{Ebert} conclude that the term is completely suppressed in the limit of very high temperature, and the Lorentz and CPT symmetries of the theory are restored. On the other hand, in \cite{N3,bpp}  it was found that the coefficient depends on the regularization schemes at zero or finite temperature, so it remains undetermined. 

It is important to recall that this problem already exists at zero temperature for massive fermions. Considering the fermionic quantum fluctuations as responsible for the induction of the CFJ model it is found that the coefficient of this term has an ambiguous value (see for instance \cite{Jackiw:1999yp, Jackiw:1999qq, Chung:1999gg, PerezVictoria:1999uh, Bonneau:2000ai, PerezVictoria:2001ej, Bonneau:2001ri, Bonneau:2006ma, Chen:2007kz}). In fact this is claimed to be an example where quantum corrections are finite but undetermined, that is, its value can only be determined through experimental inference. This is not an unusual situation in quantum field theory as was pointed out by Jackiw \cite{Jackiw:1999qq}, ocurring for instance in the Schwinger chiral model \cite{Jackiw:1984zi}. The mathematical manifestation of such phenomenon is the dependence of the result on different regularization schemes.

The main goal of the present study is to investigate whether this issue appear in the  particular case of extended massless QED at high temperature.
In the massless case at zero temperature has been shown in \cite{prdnosso} that such ambiguity is absent. It has been argued there that such result should follow naturally from dimensional projection considerations since the 4D Lorentz violating fermionic term reduces to the 3D fermionic mass term.  Since the coefficient of the 3D Maxwell-Chern-Simons model is not ambiguous \cite{Redlich}, 
the coefficient of the CFJ model for massless fermions should also be free of such pathology. This is confirmed by the results obtained previously in the Refs.\cite{Jackiw:1999yp}, \cite{PerezVictoria:1999uh} and \cite{Scarpelli:2008fw}.
For massive 4D QED this line of reasoning is not possible because the mass term has no correspondent in 3D. It is therefore suggestive that these distinct behaviors are somehow related to the breaking of the chiral symmetry. This observation prompt us to suspect that at the high temperature limit, due to the chiral symmetry restoration, the result for the massive Lorentz and CPT violating QED would be indistinguishable from that of the massless version. 
This seems to be in conflict with the fact that one cannot flow continuously from the massive to the massless limit of our theory because they have different number of degrees of freedom. However, even if the mass goes to zero or becomes negligible at high temperature, the $b$-parameter controlling the Lorentz and CPT symmetries violation still continues to play the crucial role of a residual mass in the fermion propagator structure. The $b$-parameter contrary to the mass term has the main difference of preserving the chiral symmetry and being related to the mass term of a 3D Chern-Simons theory. One of the consequence of this phenomenon, is that the function describing the Chern-Simons coefficient at finite temperature is pretty the same as that of a massive theory, as we shall see later. 
Indeed one can show that dispersion relation assumes the form:
\ben
&&\left(1-\frac{|{\bf p}|^2}{\omega^2}+\frac{M^2}{\omega^2}\right)^2-\frac{4b_0^2}{\omega^2}-\frac{4|{\bf b}|^2|{\bf p}|^2\cos^2{\theta}}{\omega^4}\nonumber\\&&-\frac{4b^2}{\omega^2}\frac{M^2}{\omega^2}+\frac{4b^4}{\omega^4}=0
\een
where $M^2=b^2-m^2$ is the ``effective mass'', such that $\frac{M^2}{\omega^2}=\frac{b^2}{\omega^2}-\frac{m^2}{\omega^2}$. The propagator
structure is maintained since $M^2\neq0$ for any value of $m^2/\omega^2\neq b^2/\omega^2$. Indeed $m^2/\omega^2\to0$ in the limit of $m^2\to0$ or $m^2\ll\omega^2$. The last regime is the limit of high temperature: $p_0\equiv\omega\sim\frac{n}{\beta}\sim{n}{T}$.

So, in this vein it seems sufficient to study the massless case to decide about the possibility of Lorentz restoration in high temperature. This is the approach we intend to take in the present work. Our main result is to show that there is no ambiguity in the induced coefficient of the Chern-Simons-like term of the CFJ model for massless fermions at high temperature and also that the breaking is more severe in this case than in the zero temperature case \cite{prdnosso} thus precluding any Lorentz symmetry restoration.

This paper is organized as follows: in section II we review the result obtained in \cite{prdnosso}, showing that there is no ambiguity in the coefficient of the Chern-Simons-like term radiatively induced in QED at zero temperature by quantum fluctuations of massless fermions. In section III we consider the same system at finite temperature and proceed to calculate the induced term by two regularization schemes, dimensional regularization and momentum cut-off regularization, explicitly showing that the coefficient of the induced term does not depend on the regularization considered as argued above. Furthermore we find that Lorentz violation is enhanced at high temperatures in this particular system, as evidenced by the fact that the coefficient is two times bigger than at zero temperature. Finally, in section IV we present our conclusions.

\section{The extended massless QED}

The model we will consider is given by the Lorentz violating model described by the following Lagrangian density \cite{12,Scarpelli:2008fw} \bea {\cal
L}=\bar{\psi}( i \pls - \bs\gamma_{5} - e \as) \psi. \label{QEDQ}
\eea In accordance with the reference \cite{prdnosso}, we are looking for the term in the effective action $S_{eff}[b,A]$, obtained by integrating the fermionic degrees of freedom, which is quadratic on the gauge field $A_{\mu}$ and of first order in derivatives. It can be expressed in the form
\bea \label{efa2a} S_{eff}^{(2)}[b,A]=\int d^4x
\Pi^{\mu\lambda\nu}\pa_{\lambda}A_{\mu}A_{\nu}, \eea with the one
loop self-energy tensor being given by \bea \label{efa2aa}
\Pi^{\mu\lambda\nu}=-\frac{ie^2}{2}\int
\frac{d^4p}{(2\pi)^4}
{\rm
tr}\left[S_b(p)\gamma^{\mu}S_b(p)\gamma^{\lambda}S_b(p)\gamma^{\nu}
\right], \eea where {\rm tr}, means that the trace just acts over
the gamma matrices and $S_{b}(p)$ is the $b_{\mu}$-dependent propagator of the theory
defined as,\bea S_b(p)=\frac{i(p+b)^{2}(\ps-\bs)}{(p^{2}-b^{2})^{2}}\,P_{L}+\frac{i(p-b)^{2}(\ps+\bs)}{(p^{2}-b^{2})^{2}}\,P_{R},
\eea where \bea P_{R,L}=\frac{1\pm\gamma_{5}}{2}. \eea
are chiral projectors. Thus, taking into account the fact that
$\{\gamma_{5},\gamma^{\mu}\}=0$ and $(\gamma_5)^2=1$ and applying the trace relation ${\rm
tr}(\gamma^{\lambda}\gamma^{\mu}\gamma^{\nu}\gamma^{\rho}\gamma_5)=4i\epsilon^{\lambda\mu\nu\rho}$,
we can write down a simple expression for the self-energy tensor
$\Pi^{\mu\nu\lambda}$: \bea\label{proj}
&&\Pi^{\mu\nu\lambda}=ie^2\,\epsilon^{\lambda\mu\nu\rho}\int\frac{d^4p}{(2\pi)^4}
\frac{1}{(p^2-b^2)^4}\times\nonumber\\&&b_{\rho}\bigl((p+b)^{4}\!+\!(p-b)^{4}\bigl)-p_{\rho}\bigl((p+b)^{4}\!-\!(p-b)^{4}\!\bigl).
\eea Now, we can write the relevant term in the effective action in the form
\bea\label{efa2aaaa} S_{eff}^{(2)}[b,A]=\frac{1}{2}\,\int d^{\,4}x
\,\epsilon^{\lambda\mu\nu\rho}k_{\rho}F_{\lambda\mu}A_{\nu}, \eea
where the $k_{\rho}$-parameter can be expressed in the form
\bea\label{tp1}
k_{\rho}&=&2ie^{2}\!\int\!\frac{d^{4}p}{(2\pi)^{4}}\Big[\frac{(p^{2}-b^{2})b_{\rho}-4(b\cdot
p)p_{\rho}}{(p^{2}-b^{2})^{3}}+\nonumber\\&&\frac{4\bigl[(p^{2}b^{2}+(b\cdot
p)^{2})b_{\rho}-2b^{2}(b\cdot
p)p_{\rho}\bigl]}{(p^{2}-b^{2})^{4}}\Big]. \eea Note that by power
counting, the momentum integral in (\ref{tp1}) involves finite terms
and terms with logarithmic divergence. However, the calculations show that the CFJ model can be induced
with its coefficient being well defined and finite in different regularization
schemes \cite{prdnosso}. The exact value is:
\bea\label{c1}
k_{\rho}=\frac{e^{2}}{8 \pi^{2}}\,b_{\rho}.
\eea

\section{The finite temperature effects}
\label{temp}
In order to analyze the effects of finite temperature in radiative correction calculations, we consider the system in a state of thermal equilibrium characterized by a temperature $T=1/\beta$. Therefore, we make the following substitutions:
\bea
p_{0}\equiv \omega_{0}=(n+1/2)\frac{2\pi}{\beta}
\,\,\,{\rm and}\,\,(1/2\pi)\int d p_{0}\rightarrow \frac{1}{\beta}\sum_{n}\eea where
$\omega_{0}$ is the Matsubara frequency for fermions. Again, we
consider the expression (\ref{tp1}) and rewrite it in the framework
of the imaginary time formalism $(x_{0}\rightarrow -ix_{0}, b_{0}\rightarrow ib_{0}, d^{4}x\rightarrow-id^{4}x, d^{4}p\rightarrow id^{4}p)$: \bea\label{T1}
k_{\rho}(\beta)&=&\frac{2i\,e^2}{\beta}\!\!\sum^{\infty}_{n=-\infty}
\!\!\int\frac{d^{3}\vec{p}}{(2\pi)^{3}}\Big[\frac{(p^{2}-
b^{2})b_{\rho}-4(b\cdot p)p_{\rho}}{(p^{2}-
b^{2})^{3}}+\nonumber\\&&\frac{4\bigl[(p^{2}b^{2}+(b\cdot
p)^{2})b_{\rho}-2b^{2}(b\cdot p)p_{\rho}\bigl]}{(p^{2}-
b^{2})^{4}}\Big]. \eea
Or, only the time-like component
\bea\label{T2.1a}
k_{0}(\beta)&=&\frac{-2i\,e^2b_{0}}{\beta}
\sum^{\infty}_{n=-\infty}\int\frac{d^{3}\vec{p}}{(2\pi)^{3}}\times\Big[\frac{3}{(\vec{p}^{2}+B^{2})^{2}}-\nonumber\\&&\frac{4\vec{p}^{2}}{(\vec{p}^{2}+B^{2})^{3}}+\frac{4b_{0}^{2}B^{2}}{(\vec{p}^{2}+B^{2})^{4}}\Big],
\eea where $B^{2}=\omega_{0}^{2}-b^{2}_{0}$.

\subsection{Dimensional regularization}

Let us now use the standard dimensional regularization scheme \cite{thooft} to perform the integration
over the spatial momentum in $D$ dimensions in the expression
(\ref{T2.1a}). The expression takes the form \bea\label{solT}
k_{0}(\beta)\!&=&\!\frac{-2ie^{2}}{(4\pi)^{D/2}}\frac{b_{0}}{\beta}\Bigl(-\frac{{a}_{0}^{2}}{b^{2}_{0}}\Bigl)^{\lambda_{1}}\!\!\sum^{\infty}_{n=-\infty}\!\!
\Bigl[\frac{\epsilon^{'}\Gamma(\lambda_{1})}{((n+\frac{1}{2})^{2}+
{a}_{0}^{2})^{\lambda_{1}}}-\nonumber\\&&\frac{{a}_{0}^{2}}{3}\frac{(3+\epsilon^{'})
\Gamma(\lambda_{2})}{((n+\frac{1}{2})^{2}+{a}_{0}^{2})^{\lambda_{2}}}\Bigl],
\eea where $\lambda_{1}=2-\frac{D}{2}$, $\lambda_{2}=3-\frac{D}{2}$,
$\epsilon^{'}=3-D$ and ${a}_{0}=ib_{0}\beta/2\pi$.

\begin{figure}[h]\centerline{\includegraphics[{angle=90,height=7.0cm,angle=-90,width=8.0cm}]
{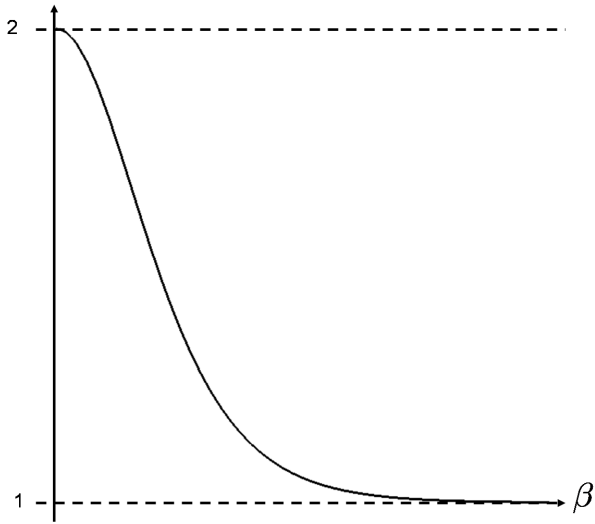}} \caption{The function $f(\beta)=1+2\pi^2 F({a}_{0})$ is different
from zero everywhere. It flows from $f(\beta)=2$ at high temperature ($\beta\to0$) to $f(\beta)=1$ at zero temperature ($\beta\to\infty$).}\label{fig1}
\end{figure}

At this point we need an explicit expression for the sum over the
Matsubara frequencies. We use the following result \cite{FO}
\bea\label{fo} &&\sum_n \frac{1}{[(n+\frac{1}{2})^2 +
{a}_{0}^{2}]^{\lambda}}=\frac{\sqrt{\pi}\Gamma(\lambda -
1/2)}{\Gamma(\lambda)({a}_{0}^{2})^{\lambda - 1/2}}
+\nonumber\\
&&\!\!\!\!4\sin(\pi\lambda)\!\!\int_{|{a}_{0}|}^\infty \!\!\frac{dz}{(z^2 -
{a}_{0}^{2})^{\lambda}} Re\!\!\left(\frac{1}{\exp 2\pi(z +
\frac{i}{2}) -1}\!\!\right)\!, \eea which is valid for $1/2<\lambda<1$.
Substituting this formula into (\ref{solT}) we have \bea\label{T3}
&&k_{0}(\beta)=\frac{2ie^{2}}{(4\pi)^{D/2}}\frac{b_{0}}{\beta}\Bigl(-\frac{{a}_{0}^{2}}{b^{2}_{0}}\Bigl)^{\lambda_{1}}\Bigl[\frac{\sqrt{\pi}\Gamma(1+\frac{\epsilon{'}}{2})}{({a}_{0})^{\epsilon{'}}}\Bigl(1-\frac{\epsilon{'}}{3}\Bigl)+\nonumber\\&&4\epsilon{'}\Gamma(\lambda_{1})\sin(\pi\lambda_{1})\int_{|{a}_{0}|}^\infty
\frac{dz}{(z^2 - {a}_{0}^{2})^{\lambda_{1}}}
\times\nonumber\\&&Re\Bigl(\frac{1}{\exp 2\pi(z + \frac{i}{2}) -1}\Bigl)-\;\;\frac{4
(3+\epsilon{'}){a}_{0}^{2}}{3}\Gamma(\lambda_{2})\times\nonumber\\&&\!\!\sin(\pi\lambda_{2})\!\!\int_{|{a}_{0}|}^\infty
\!\frac{dz}{(z^2 - {a}_{0}^{2})^{\lambda_{2}}}
Re\Bigl(\frac{1}{\exp 2\pi(z + \frac{i}{2}) -1}\Bigl)\Bigl]. \eea
Note that for $\lambda_1=2-\frac{D}{2}$ and
$\lambda_2=3-\frac{D}{2}$ we cannot apply this relation for $D=3$
since the integral diverges. Thus, we carry out the analytic
continuation for this relation, so that we obtain \bea
\label{ancont} &&\int_{|{a}_{0}|}^\infty \frac{dz}{(z^2 -
{a}_{0}^{2})^{\lambda}}
Re\Big(\frac{1}{\exp 2\pi(z + ib) -1}\Big)=
\frac{1}{2 {a}_{0}^{2}}\times\nonumber\\&&\frac{3-2\lambda}{1-\lambda}
\int_{|{a}_{0}|}^\infty \frac{dz}{(z^2 -
{a}_{0}^{2})^{\lambda-1}}Re\Big(\frac{1}{\exp 2\pi(z + ib)
-1}\Big)-\nonumber\\&&\frac{1}{4
{a}_{0}^{2}}\frac{1}{(2-\lambda)(1-\lambda)}
\int_{|{a}_{0}|}^\infty \frac{dz}{(z^2 -
{a}_{0}^{2})^{\lambda-2}}
\times\nonumber\\&&\frac{d^2}{dz^2}Re\Big(\frac{1}{\exp 2\pi(z + ib)
-1}\Big).\eea Now we can substitute this expression into
(\ref{T3}) and use it for $D=3$. Thus, after some simplifications we
get \bea\label{fdp04}
k_{0}(\beta)=\frac{ie^{2}}{8\pi^{2}}[1+2\pi^{2}F({a}_{0})]\,b_{0},
\eea where the function $F({a}_{0})$ is given by
\be\label{fp4}
F(a_{0})=\int_{|{a}_{0}|}^{\infty}dz(z^2-{a}_{0}^{2})^{1/2}
\frac{\tanh(\pi z)}{\cosh^2(\pi z)}, \ee and has the following
limits: $F({a}_{0}\to\infty)\to0$ ($T\to0$) and
$F({a}_{0}\to0)\to 1/2\pi^{2}$ ($T\to\infty$) --- see
Fig.\ref{fig1}.

In summary, we conclude that the Chern--Simons-like coefficient at
finite temperature regularized via dimensional regularization scheme
is given by \bea &&k_{0}=\frac{ie^2}{4\pi^2}b_{0} \;\;\;\;
({a}_{0}\to0 \mbox{ or } T\to\infty), \eea and \bea
\;\;\;\;\;\;\;k_{0}=\frac{ie^2}{8\pi^2}b_{0}\;\;\;\;({a}_{0}\to\infty\mbox{
or } T\to0). \eea Notice that the coefficient $k_{0}$ at $T\to0$
coincides with the coefficient (\ref{c1})
previously obtained  at zero temperature, while  at $T\to\infty$ it
corresponds to the double of this value.
A similar result was found in \cite{10} for a {\it
massive} theory by using the dimensional and momentum cut-off
regularization. In this particular case the {\it massless} theory at
high temperature leads to the same result of a massive theory
without temperature.

\subsection{Momentum cut-off-$\Lambda$ regularization}
To develop calculations via momentum cut-off, we rewrite the expression (\ref{T2.1a})
in spherical coordinates, and as result, we obtain
\bea\label{c11}
k_{0}(\beta)&=&\frac{-ie^{2}\,b_{0}}{\beta \pi^{2}}\sum^{\infty}_{n=-\infty}\int^{\infty}_{0}
d\vec{p}\,\vec{p}^{2}\Big[\frac{3}{(\vec{p}^{2}+B^{2})^{2}}-\nonumber\\&&\frac{4\vec{p}^{2}}{(\vec{p}^{2}+B^{2})^{3}}+\frac{4b_{0}^{2}B^{2}}{(\vec{p}^{2}+B^{2})^{4}}\Big]
\nonumber\\&=&\frac{-ie^{2}b_{0}}{\beta \pi^{2}}\sum^{\infty}_{n=-\infty}
\frac{\Big[I_{1}(\Lambda)+I_{2}(\Lambda)+I_{3}(\Lambda)\Big]}{B}
\eea
where the integrals $I_{1}(\Lambda)$, $I_{2}(\Lambda)$ and $I_{3}(\Lambda)$ are written as,
\bea
I_{1}(\Lambda)&=&\int^{u=\Lambda/B}_{0} du \frac{3 u^{2}}{(u^{2}+1)^{2}}\nonumber\\
&=&\frac{3}{2}\frac{\Big[\arctan(\tilde{\Lambda})+\arctan(\tilde{\Lambda})\tilde{B}^{2}-\tilde{B}\Big]}{(1+\tilde{B}^{2})},
\eea

\bea
I_{2}(\Lambda)&=&-\int^{u=\Lambda/B}_{0} du \frac{4 u^{4}}{(u^{2}+1)^{3}}\nonumber\\
&=&\frac{-1}{2(1+\tilde{B}^{2})^{2}}\Big[3\arctan(\tilde{\Lambda})+6\tilde{B}^{2}\arctan(\tilde{\Lambda})+\nonumber\\&&3\tilde{B}^{4}\arctan(\tilde{\Lambda})-5\tilde{B}-3\tilde{B}^{3}\Big]
\eea and \bea I_{3}(\Lambda)&=&\int^{u=\Lambda/B}_{0} du \frac{4
b^{2}_{0}u^{2}}{(u^{2}+1)^{4}}\nonumber\\&=& \frac{b^{2}}{12
B^{2}}\frac{1}{(1+3\tilde{B}^{2}+3\tilde{B}^{4}+\tilde{B}^{6})}\Big[3\arctan(\tilde{\Lambda})
+\nonumber\\&&9\tilde{B}^{2}\arctan(\tilde{\Lambda})+9\tilde{B}^{4}\arctan(\tilde{\Lambda})+\nonumber\\&
&3\tilde{B}^{6}\arctan(\tilde{\Lambda})+3\tilde{B}+8\tilde{B}^{3}-3\tilde{B}^{5}\Big]
\eea with $\tilde{\Lambda}=\Lambda/B$ and $\tilde{B}=B/\Lambda$.
Therefore, taking the limit $\Lambda\rightarrow\infty$, this implies
in $\tilde{\Lambda}=\infty$ and $\tilde{B}=0$. Thus, we have the
following results: \bea I_{1}(\Lambda)=\frac{3\pi}{4}\;\;\;,\;
I_{2}(\Lambda)=-\frac{3\pi}{4}\;\;\;{\rm
and}\;I_{3}(\Lambda)=\frac{\pi b^{2}_{0}}{8 B^{2}}. \eea Now, we can
substitute the values of the integrals above in expression
(\ref{c1}) and obtain the simple equation \bea\label{c2}
k_{0}(\beta)=\frac{ie^{2}}{16\pi^{2}}\sum^{\infty}_{n=-\infty}\frac{a^{2}_{0}}{((n+\frac{1}{2})^{2}
+a_{0}^{2})^{\frac{3}{2}}}.
\eea Now, we shall perform the analytical continuation in the
relation (\ref{fo}) and substitute the full result of sum in
(\ref{c2}) for $\lambda=\frac{3}{2}$. In this way, we must get
exactly the expression (\ref{fdp04}) in the dimensional
regularization scheme with $k_{0}=ie^{2}b_{0}/8\pi^{2}$ for the
regime of zero temperature and $k_{0}=ie^{2}b_{0}/4\pi^{2}$ for
high temperature.
\\


\section{Conclusions}
\label{conc}

In this paper we have extended our recent results in
Lorentz-violating massless QED to the finite temperature environment. We had
previously shown that contrary to the massive case, in massless
extensions of Lorentz-violating QED at zero temperature, it
seems that the 3+1 dimensional Chern--Simons-like term is finite and
determined because, by dimensional reduction, one can see that at
least one of its components is related to its finite 2+1 dimensional
Chern--Simons counterpart. In order to support this conjecture we
have used different regularization schemes to show that we find the
same answer. In the present investigation, we have shown that this
continues to be true at finite temperature. In the limit of high
temperatures we find the same answer for the coefficient of the 3+1
dimensional Chern--Simons via two distinct regularization schemes: dimensional regularization and the momentum cut-off regularization.
Furthermore, we observe that the Chern--Simons-like coefficient
at high temperature is two times bigger than at zero
temperature. That is, of course, a $nonzero$ result contrary to what
one could expect if one believes that high temperature is always
bounded to restore symmetries --- at least in part \cite{bazeia}. However, a few comments concerning
this fact are in order. The same phenomenon has already appeared in the massive case \cite{N3,bpp}. And in fact there is no evidence to believe that any symmetry should be restored at high temperature. There are other examples in Nature of systems, such as Rochelle salts, where a symmetry is broken at high temperature and restored at low temperature \cite{ns}. What we have shown is an example of a system where the Lorentz symmetry is broken at zero temperature and continues to be broken at high temperature with an even higher strength as determined by the Chern--Simons-like coefficient. 

{\acknowledgments} We would like to thank CNPq,
CAPES, PNPD/PRO\-CAD-CAPES, and FAPERJ for partial financial support. C. Wotzasek and F.A. Brito also acknowledge D. Bazeia for interesting discussions.

\end{document}